\documentclass[aps,prb,twocolumn,floatfix]{revtex4}

\usepackage{graphicx}
\usepackage{amsmath}
\usepackage[T1]{fontenc}
\usepackage[latin1]{inputenc}
\usepackage{times}
\usepackage[normalem]{ulem}

  \makeatletter
  \def\@dotsep{4.5}
  \makeatother

\newlength{\myVSpace}
\newcommand{\ket}[1]{\left| #1 \right\rangle}
\newcommand{\bra}[1]{\left\langle #1 \right|}

\begin{document}

\title{Effect of matrix parameters on mesoporous matrix based quantum computation}

\author{T.E. Hodgson}
\affiliation{Department of Physics, University of York, Heslington, York,
YO10 5DD, United Kingdom}
\author{M.F. Bertino}
\affiliation{Department of Physics, University of Missouri-Rolla, Rolla, MO
65409, USA}
\author{N. Leventis}
\affiliation{Department of Chemistry, University of Missouri-Rolla, Rolla, MO
65409, USA}
\author{I. D'Amico}
\affiliation{Department of Physics, University of York, Heslington, York,
YO10 5DD, United Kingdom.}

\begin{abstract}
We present a solid state implementation of quantum computation, which improves previously proposed optically driven schemes. Our proposal is based on vertical arrays of quantum dots embedded in a mesoporous material which can be fabricated with present technology. We study the feasibility of performing quantum computation with different mesoporous matrices. We analyse which matrix materials ensure that each individual stack of quantum dots can be considered isolated from the rest of the ensemble-a key requirement of our scheme. This requirement is satisfied for all matrix materials for feasible structure parameters and GaN/AlN based quantum dots. We also show that one dimensional ensembles substantially improve performances, even of CdSe/CdS based quantum dots.
\end{abstract}

\maketitle

\section{Introduction} 
Solid state based hardware is considered a promising candidate for implementing quantum computing\cite{neilsenchang}. This is due to the potential scalability of the involved nanostructures, and to the relative ease of integrating them with more traditional electronic components, e.g. for input/output functions. Some interesting proposals rely on spin and exciton qubits confined in semiconductor self-assembled quantum dots (QD's)\cite{PhysToday}, which can be manipulated using ultra fast laser pulses\cite{chapter,brendon,sham}. In these schemes qubit coupling, which is necessary to perform universal quantum computation, can be realised through exciton-exciton direct Coulomb interaction\cite{irene,gan,paulibl}. The presence of an exciton in a QD produces a biexcitonic shift in the ground state excitonic energy of a nearby QD. By driving a qubit at this shifted frequency, conditional operations can be performed\cite{irene,gan}.

Currently two major problems for QD based quantum computation are addressability of the single qubit and size polydispersity in self-assembled QD ensembles. In order to carry out practical quantum computation it is necessary to be able to address individual qubits. This poses a problem for optically driving the response of self assembled quantum dots (such as the ones grown by Stranski-Krastanow techniques): in these dot ensembles the size of each dot is one order of magnitude smaller than that of the laser spot addressing it. The use of energy selective methods on isolated stacks of quantum dots (quantum registers) has been proposed to solve this problem\cite{irene,gan}. It is still experimentally difficult to control QD size and position in a satisfactory way, and QD vertical stacks tend to form at random positions. Additionally the size of the QDs within the stacks is mainly due to strain propagation and it is hardly controllable, resulting in the practical difficulty of creating the desired sequences of energy selectable excitonic transitions. To overcome these difficulties we recently proposed to use mesoporous matrices to create regular arrays of QD qubits suitable for quantum computation\cite{mesoporous}. The redundant encoding typical of the chosen hardware protects the computation against gate errors and the effects of measurement induced noise. We have discussed that quantum computation can be efficiently implemented using this hardware for both III-V and II-VI materials in a TiO$_2$ matrix. In this paper we analyse how the matrix material affects the implementation of quantum computation.

\section{Computational hardware} 

%%% 2.1 %%%

We consider alternating layers of two semiconductors with widely different band gaps deposited in a mesoporous matrix. This provides a stack of QDs (qubits), sandwiched between the larger band gap material (barriers). The resulting system is depicted in Fig.~1(a) and consists of an array of identical, hexagonally-packed stacks of quantum dots (columns). The band structure within each column is sketched in Fig.~1(b). 

We consider semiconductors which assume wurtzite crystal structure for both the barriers and qubits. This leads to strong intrinsic electric fields\cite{piezo} within each column, (see Fig.~1(b)). The built-in electric field enhances coupling between excitons in neighbouring quantum dots {\it within a stack}\cite{gan}. Quantum computation can then be carried out by using sequences of laser pulses, as described in Ref.~\cite{chapter}. Under the influence of the same laser pulse, each column within the laser spot will act as an independent replica of the same computational array. This intrinsic large redundance protects the computation even in the presence of errors and measurement induced noise\cite{mesoporous}.

\begin{figure}%f1
  \includegraphics*[width=\linewidth]{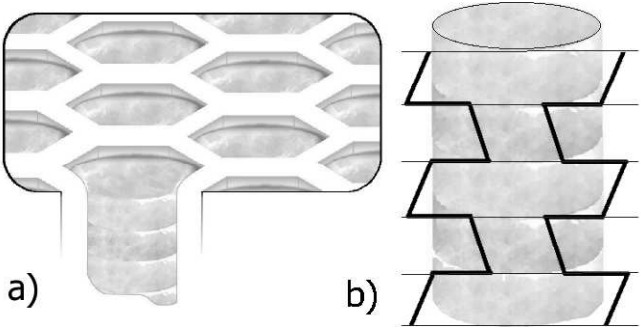}
  \caption{(a) The proposed system, an ensemble of stacks of alternating QDs and barriers embedded in a mesoporous matrix. (b) The band structure of each individual stack including the intrinsic field of the materials.}
\end{figure}

\section{Theoretical Model}
Each individual stack is modelled as a column of cylindrical quantum dots with the same radius. To calculate the biexcitonic shifts, the confining potentials are approximated as parabolic potentials with the same characteristic widths as the expectation values $\sqrt{<z^2>}$ and $\sqrt{<r^2>}$ of the cylindrical dots in the stack.  Nearby quantum dots (qubits) are coupled by the biexcitonic shift $\Delta E $ between ground state excitons. In order to calculate interaction energies between qubits we approximate the biexcitonic shift $\Delta E$ as
\begin{equation}
| \Delta E|=|\bra{\psi_1\psi_2}U_C\ket{\psi_1\psi_2}|,\label{integral}
\end{equation}
where $\psi_i(r_{ie},r_{ih})=\psi_{ie}(r_{ie})\psi_{ih}(r_{ih})$ is the wavefunction of the exciton in QD$_i$ in the single particle approximation, and $U_C$ is the Coulomb interaction between the two excitons. In order for the scheme to be feasible two conditions must be satisfied. The biexcitonic shift between neighbouring qubits within a stack must be large enough, so quantum computation can be performed within the stack on time scales much shorter than the relevant decoherence times, and the inter-stack interactions must be negligable so that each QD array can be considered isolated.

The biexcitonic shift between excitons within a stack is exploited to perform two-qubit gates using multicolour trains of laser pulses\cite{irene,gan,paulibl}. For performing operations on picosecond time scales -- which is essential due to the relatively short excitonic decoherence times --  $\Delta E $  must be of the order of a few meV. When choosing the correct parameter range however, additional factors must be taken into consideration\cite{irene}.  Larger $\Delta E $ can be induced by increasing the height of the quantum dots. This increases the excitonic dipole moments under the influence of the intrinsic electric field but reduces the oscillator strength, i.e. the efficiency of driving the computation through coupling to laser pulses. Finally the barrier width must be large enough to ensure that single particle tunnelling between quantum dots is negligible on the relevant time scales. 

If all the stacks in the ensemble were to compute correctly -- i.e. no computational errors -- inter-stack interactions would only renormalise the excitonic energies by the same amount for each stack, which could be easily accounted for in the computational scheme. However the event of computational failure in a certain stack would induce a local, unwanted, shift  of the exciton energies in neighbouring stacks. Each individual stack will interact with a certain number of such 'faulty computing' stacks. Therefore the magnitude of this unwanted shift is the sum of the interaction energies of a resident exciton with an exciton in each 'faulty computing' stack $|\Delta E_{inter}^{tot}|$.  The inter-stack interaction between qubits is negligible, if it is much smaller than the interaction between excitons within a column,  i.e. $|\Delta E_{inter}^{tot}|<<|\Delta E|$.
 The stacks are arranged in a hexagonal structure, which can be represented as a series of concentric hexagonal shells surrounding each stack. The $i$~th shell consists of $6i$ stacks. The number of expected computational failures in the $i$~th shell will therefore be $6i\cdot(1-p)$, where $p$ is the probability of a successful computation for any individual stack. We estimate the total energy shift due to interactions with all the `faulty computing' stacks in the ensemble as:
\begin{equation}
|\Delta E_{inter}^{tot}|=\sum_{i=1}^{ensemble} 6i(1-p)\Delta E_{inter}(r_i) \label{totalinter}
\end{equation}
where $r_i=(r_{max}+r_{min})/2$ is the average radius of the $i$~th hexagonal shell, $r_{max}$ and $r_{min}$ being the maximum and minimum distance of the shell from the central stack, and $\Delta E_{inter}$ is calculated according to Eq.~(\ref{integral}) considering dots in different stacks.

\section{Influence of material parameters and dimensionality}
We have shown\cite{mesoporous} that for typical III-V and II-VI materials embedded in a  TiO$_2$ matrix, the computational hardware can be designed such that the biexcitonic shift between qubits within a stack is large enough to perform quantum computation, and that the inter-stack interactions are small enough for the hardware to be considered an ensemble of isolated stacks. For a GaN QD and a AlN barrier, or for a CdSe QD and a CdS barrier, each of radius 5nm, there exists a range of QD heights and barrier widths which lead to a sufficient biexcitonic shift between neighbouring qubits ($\Delta E > 3meV$), satisfactory oscillator strength ($\mu>0.15\mu_{zerofield}$), and a sufficiently long tunnelling time between qubits ($\tau>1ns$). The tunnelling time is estimated by taking the inverse of the tunnelling rate as expressed in Ref.~\cite{storage}.
\begin{figure}%f1
  \includegraphics*[width=\linewidth]{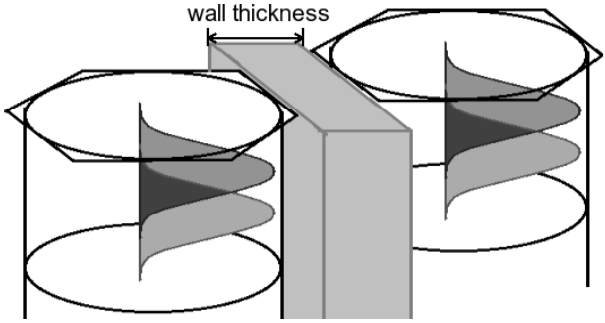}
  \caption{Interstack qubits separated by the matrix wall.}
\end{figure}

 The interaction between neighbouring stacks can be decreased by increasing the thickness of the matrix walls (see Fig. 2.), which can be made up to $\approx6$~nm thick before the order of the array is lost. The magnitude of the interaction with faulty computing stacks depends also on the probability $p$ of each stack computing successfully (see Eq (2)). It was found\cite{mesoporous} that for the GaN/AlN system with TiO$_2$ matrix, and matrix wall thicknesses of $6$nm, each stack could be considered isolated from the rest of the ensemble for all $p$. For a CdSe/CdS hardware it was found that for matrix walls $6$~nm thick the stacks could be considered isolated for $p\stackrel{>}{\sim}0.86$, and similarly for a wall thickness of $4.5$~nm for $p\stackrel{>}{\sim}0.91$.

\begin{figure}%f1
  \includegraphics*[width=\linewidth]{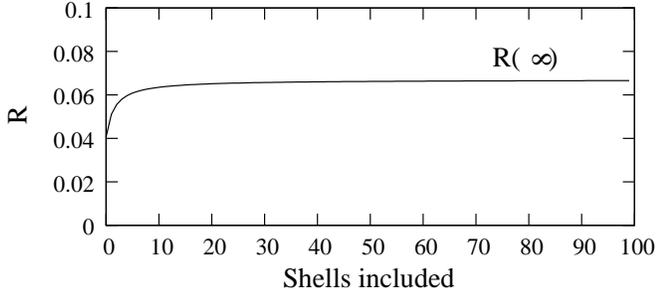}
  \caption{$R$ versus the number of shells included in the calulation, for the GaN/AlN system in a TiO$_2$ matrix with $2$~nm walls, and $p=0.83$}
\end{figure}

Figure 3 shows the ratio $R=|\Delta E^{tot}_{inter}|/|\Delta E|$ against the number of shells of neighbours included in the calculation, for the GaN/AlN system in a TiO$_2$ matrix ($\epsilon_{TiO_2} \approx 100$)\cite{TiO2} with $2$~nm walls, and $p=0.83$. $R(\infty)$ is the number at which this ratio saturates. For the interstack interaction to be negligable it must be an order of magnitude smaller than $\Delta E$ i.e. $R(\infty)<0.1$. For the case in Fig.3. $R(\infty)\approx0.066$, therefore the interstack interactions can be neglected and each stack in the ensemble can be considered isolated. 

TiO$_2$ was initially proposed as the matrix material due to its high dielectric constant, resulting in high screening between stacks. However experimentally other materials may be used to produce mesoporous matrices. In the following we will analyse how different materials, characterised by different dielectric constants, would affect the feasibility of our scheme by increasing or reducing the inter-stack interaction. Fig. 4. shows $R(\infty)$ for a GaN/AlN system against the dielectric constant of the matrix for different $p$ for matrix walls a) $2$~nm thick and b) $6$~nm thick. It can be seen that the condition $R(\infty)<0.1$ still holds for materials with dielectric constants much lower than TiO$_2$. For example a much more standard material for the matrix is SiO$_2$ which has a dielectric constant $\epsilon\approx4$. Fig.4. shows that for a SiO$_2$ matrix with walls as thin as $2$~nm and $p\stackrel{>}{\sim}0.9$ the inter-stack interaction is already negligable. For $6$~nm walls SiO$_2$ is a suitable matrix for $p\stackrel{>}{\sim}0.83$. We underline that $p=0.83$ corresponds to, on average, one nearest neighbour failing. The situation is different for a CdSe/CdS based system. As shown in Fig. 5, when $p=0.83$, materials with dielectric constant $\epsilon>130$ would be needed to achieve an ensemble of isolated stacks. 

\begin{figure}%f1
  \includegraphics*[width=\linewidth]{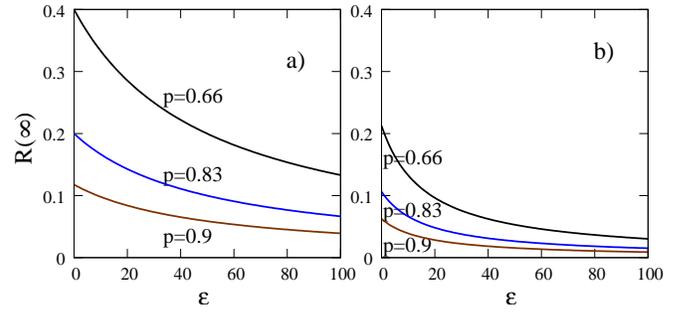}
  \caption{$R(\infty)$ for a GaN/AlN system versus the dielectric constant of the matrix for different $p$. a) matrix walls $2$~nm thick and b) matrix walls $6$~nm thick matrix walls.}
\end{figure}

\begin{figure}%f1
  \includegraphics*[width=\linewidth]{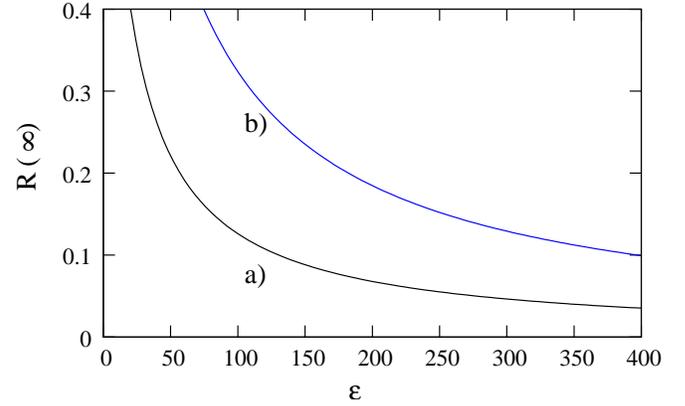}
  \caption{$R(\infty)$ for a CdSe/CdS system versus the dielectric constant of the matrix for $p=0.83$, for a) matrix walls $6$~nm thick and b) $3$~nm thick matrix walls.}
\end{figure}

One possible solution to decrease the interaction between stacks for the II-VI system would be to reduce the dimensionality of the ensemble array. By reducing the 2D array of stacks to a 1D chain the interaction energy between a single stack and the 'faulty computing' stacks in the ensemble is reduced to
\begin{equation}
|\Delta E_{inter}^{tot}|=2(1-p)\sum_{i=1}^{ensemble}\Delta E_{inter}(r_i) \label{1d}.
\end{equation}
Figure 6 shows the ratio $R$, against the number of neighbours included in the calculation for a TiO$_2$ matrix and a chain of CdSe/CdS-based stacks. It is shown that for a wall thickness of $3$~nm the chain can be considered isolated for $p$ as low as $0.76$ (and for even lower $p$ when the walls are thicker). This is a large improvement in respect to the results for the 2D array shown in Fig. 5b. Reducing the dimensionality of the array has not been experimentally achieved. If the conditions during manufacture could be changed to produce a rectangular lattice instead of hexagonal (this change alone would reduce the interaction energy between stacks), an effective sytem of 1D chains could be realised by having one lattice constant much greater than the other. Such a configuration in fact could be seen as a group of independent chains of stacks. 

\begin{figure}%f1
  \includegraphics*[width=\linewidth]{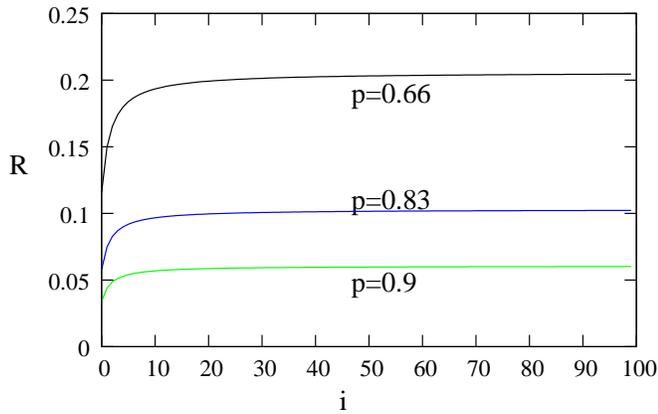}
  \caption{$R$ for a chain of CdSe/CdS stacks, against the number of pairs of nearest neighbours included in the calculation for $p=0.83$, and for $3$~nm thick matrix walls.}
\end{figure}

\section{Conclusions}
Our proposal is based on a mesoporous matrix which provides an uncorrelated ensemble of computational arrays. This intrinsic redundancy improves the response of the system, when used as quantum computation hardware\cite{mesoporous}. We have discussed how the use of different matrix materials would influence this property, namely by increasing/decreasing inter-stack interactions. It has been shown that, for GaN-based qubits, the stacks in the ensemble can be considered isolated for any reasonable matrix material, provided that the matrix walls are $6$~nm thick and the probability of each isolated stack in the ensemble computing successfully is greater than a $0.83$. The stacks in the ensemble can still be considered isolated for matrices with thinner walls as long as there is a higher probability that each individual stack will compute correctly. It is noteworthy mentioning that, when $p\stackrel{>}{\sim}0.9$, even a material with a relatively low dielectric constant such as SiO$_2$ could be used, and with matrix walls as thin as $2$~nm.

Finally we have shown that reducing the dimensionality of the ensemble also improves the response of the system, by reducing the influence of interactions between the members of the ensemble. This is particularly important for CdSe/CdS-based systems. In this case, for a $TiO_2$ matrix with hexagonally packed pores, to be able to regard the QD stacks as independent, the matrix walls must be at least $4.5$~nm thick and the probability of individual success as high as $p=0.9$\cite{mesoporous}. However the use of matrices with pores aligned along 1D chains could allow the probability of success for isolated stacks to be as low as $0.76$, with matrix walls as thin as $3$~nm.

\end{document}